\documentclass[10pt]{article}
\usepackage{amsfonts}
\usepackage{graphicx}
\usepackage{epsfig}
\usepackage{color}

\def\beq{\begin{equation}}
\def\eeq{\end{equation}}
\def\beqa{\begin{eqnarray}}
\def\eeqa{\end{eqnarray}}

\def\hf{\textstyle{1\over2}}

\newcommand{\sucg}[6]{\left\langle {#1 \atop #2} \ {#3 \atop #4}\,\right\vert\, \left.{#5\atop #6}\right\rangle }
\newcommand{\surcg}[6]{\left\langle {#1 \atop #2} \ {#3 \atop #4}\,\right\Vert\, \left.{#5\atop #6}\right\rangle }

\newcommand{\Su}[1]{$\mathfrak{SU}(#1)$}

\def\u3{$\mathfrak{u}(3)$}
\def\U3{$\mathfrak{U}(3)$}
\def\so3{$\mathfrak{so}(3)$}
\def\So3{$\mathfrak{SO}(3)$}

\begin{document}

\title{Quantum Tomography of Three--Level Atoms}
\author{Andrei B. Klimov $^1$ and Hubert de Guise $^2$ \\
$^1$ Departamento de F\'{\i}sica, Universidad de Guadalajara,
\\ 44420
Guadalajara, Jalisco, Mexico\\
$^2$ Department of Physics, Lakehead University, \\Thunder Bay,
Ontario P7B 5E1, Canada}

\date{November 2007}

\maketitle

\begin{abstract}
We analyze the possibility of tomographic reconstruction of a system
of three-level atoms in both non-degenerate and degenerate cases. In
the non-degenerate case (when both transitions can be accessed
independently) a complete reconstruction is possible. In the
degenerate case (when both transitions are excited simultaneously)
the complete reconstruction is achievable only for a single atom in
the $\Xi $ configuration.  For multiple $\Xi$ atoms, or even a
single atom in the $\Lambda$ configuration, only partial
reconstruction is possible. Examples of one and two-atom cases are
explicitly considered.
\end{abstract}

\section{Introduction.}

Quantum tomography is now an accepted technique to reconstruct the
state of a quantum system \cite{2}. Recent applications include the
reconstruction of numerous physical systems, such as a radiation
field \cite{3}, trapped ions and molecular vibrational states
\cite{4}-\cite{8}, spin \cite{11}\cite{16} and some other systems
\cite{19}\cite{Karassiov}.

The main idea behind Quantum Tomography is to use population measurements of
the ``rotated'' density matrix of the system. Explicitly, if $\rho$ is the
density matrix, a tomogram $\omega(\psi,\kappa)$ is defined by
\begin{equation}
\omega (\psi ,\kappa )\equiv\left\langle \psi \right|U\left( \kappa
\right)\, \rho \,U^{-1}\left( \kappa \right)\left| \psi \right\rangle
\label{tomogram}
\end{equation}
where
\begin{equation}
\rho(\kappa)\equiv U\left( \kappa \right) \rho \,U^{-1}\left( \kappa \right)
\end{equation}
is the density matrix rotated by the unitary transformation $U$ and $\kappa $
stands for all the parameters required to uniquely specify $U$.
By varying the input parameters $\kappa$, one obtains a complete collection
of different observables (called a \emph{quorum}), from which a
characterization of the initial quantum state of the system can be obtained
\cite{12}.

Every tomographic scheme is centered on the possibility of inverting Eq.(\ref%
{tomogram}). This inversion and the corresponding reconstruction of $\rho $
from $\omega (\psi ,\kappa )$ can always be mathematically achieved \cite%
{BrifMann} \cite{Klimov} when a Lie group $\mathfrak{G}$ acts
irreducibly in the Hilbert space appropriate for the description of
the quantum system under investigation. Experimental success depends
on whether or not all the requirements element in $\mathfrak{G}$ can
be practically implemented.

The objective of this paper is to study the possible tomographic
reconstruction of the density matrix describing a collection of three--level
atoms. Several approaches have already been proposed to reconstruct the
state of a general three--level system (qutrit). In \cite{15}\cite{qutritrec}%
\cite{qutrit}, the reconstruction of the quantum state of a three--level
optical system is implemented for a frequency- and estimating quantum states
and measuring fourth--order field moments. The use of non-orthogonal
measurements as a way to reconstruct the state of a system (provided those
measurements span the Hilbert space) as well as a detailed example of
reconstruction for one and two--qutrit systems, is considered in \cite{16}.

Even though one technically may require only a finite number of
different experimental set-ups for complete tomographic
reconstruction of atomic states, we will focus on the so-called
redundant reconstruction, which implies a continuous set of
measurements ``blanketing" all the parameter space. We justify this
redundancy on the grounds that the reconstruction of a many-atom
system would, in practice, require a large number of such discrete
measurements.

Our strategy is to investigate tomographic reconstruction of the atomic
state by probing atoms through the application of a carefully selected
sequence of dispersive and resonant electromagnetic pulses.

Once an initial pulse of classical light has created a state of collective
excitation in an ensemble of cold atoms, another pulse converts the atomic
excitations into field excitations generating, for different atomic
configurations, photons in a well-defined spatial and temporal mode \cite%
{singlephotongenerate}. The total number of such photons is
determined by photoelectric detection so that a probability of
detecting a given number of photons can be directly related to a
tomogram \cite{ross}.

In this work, our analysis will focus on two of three
fundamentally different atomic configurations. They are the so--called $%
\Lambda$ and $\Xi$ configurations, distinguished by the presence of
transition degeneracies from the generic \emph{non--degenerate}
configuration.

In Sec.\ref{nondegenerate}, we discuss this non--degenerate case,
where transition frequencies are essentially different for distinct
atomic transitions and where each transition can be independently
interrogated by a pulse of the appropriate frequency. In this case,
the density matrix can be completely reconstructed.

In Sec.\ref{sec:lambdacase} and \ref{sec:xicase}, the systems under
consideration contain atomic transitions having the same frequency; it is
\emph{not} possible to interrogate every transition individually. We will
consider these cases at length and highlight the differences between the
(global) symmetries pertinent to the description of these inequivalent
degenerate atomic configuration. We will show, for these cases, that the
density matrix cannot in general be completely reconstructed and that the
partial information extracted from the measurements in the case of $\Lambda $
and $\Xi $ atoms is essentially different.


\section{\u3, $\mathfrak{su}(3 )$ and three--level atoms}

\label{U3}

The Hamiltonian governing the evolution of a collection of $A$ three--level
atoms in a classical field has the form
\begin{equation}
\hat{H}=\hat{H}_{0}+\hat{H}_{12}+\hat{H}_{23}  \label{hamiltoniangen}
\end{equation}
where $\hat{H}_{0}$ is the free atomic Hamiltonian, and $\hat{H}_{ij}$ is
the interaction term between levels $i$ and $j$.

The terms in Eq.(\ref{hamiltoniangen}) are most transparently analyzed by
introducing a set $\{\hat{S}_{ij};i,j=1,2,3\}$ of collective transition
operators that satisfy the standard commutation relations of the \u3
algebra:
\begin{equation}
[ \hat{S}_{ij},\hat{S}_{kl}] =\delta _{jk}\hat{S}_{il}-\delta _{il}\hat{S}%
_{kj}\, .
\end{equation}
Thus, the Hilbert space for our systems naturally decomposes into a
sum of subspaces invariant under the action of the Lie group \U3.

We further assume that the $A$ atoms are indistinguishable and their
states are fully symmetric under permutation of the particle
indices. Hence, the possible states of our system belong to a single
unitariy irreducible representation of \U3 having dimension
$\hf(A+1)(A+2)$ and denoted by $(A,0)$ in mathematics. The dimension
of the Hilbert space is given by the number of ways of distributing
$A$ bosons in three modes.

If we introduce the atomic basis
\begin{equation}
\{\vert n_1n_2n_3 \rangle\, ,n_1,n_2,n_3\ge 0\, ,n_1+n_2+n_3=A\}\, ,
\end{equation}
where ${n_{j}}$ denotes the population in the $j$-th atomic level, the
matrix elements of $\hat{S}_{ij}$ can be easily evaluated using the
Schwinger realization:
\begin{equation}
\hat{S}_{ij}\mapsto a_{i}^{\dagger }\,a_{j}\, .  \label{Schwinger}
\end{equation}
In the one--atom case, this yields
\begin{equation}
\hat{S}_{ij}|j\rangle =|i\rangle ,
\end{equation}%
where the following identification
\begin{equation}
|100\rangle \leftrightarrow |1\rangle ,\quad |010\rangle \leftrightarrow
|2\rangle \,,\quad |001\rangle \leftrightarrow |3\rangle \,,
\end{equation}%
has been made. Throughout this work, we will assume the ordering $E_{1}\le
E_{2}\le E_{3}$ of individual atomic levels.

In terms of $\hat{S}_{ij}$'s, the Hamiltonian of Eq.(\ref{hamiltoniangen})
takes the form
\begin{eqnarray}
\hat{H}_{0} &=&\displaystyle\sum_{i=1}^{3}\,E_{i}\,\hat{S}_{ii}\,, \\
\hat{H}_{12} &=&g_{1}\left( e^{i\omega _{1}t}\hat{S}_{12}+e^{-i\omega _{1}t}%
\hat{S}_{21}\right) \,,\quad  \label{h23} \\
\hat{H}_{23} &=&g_{2}\left( e^{i\omega _{2}t}\hat{S}_{23}+e^{-i\omega _{2}t}%
\hat{S}_{32}\right) ,  \label{h12}
\end{eqnarray}%
where $\omega _{1}$ and $\omega _{2}$ are frequencies of the external fields
and $g_{1}$ and $g_{2}$ are coupling constants, chosen to be real for
simplicity.

The operator
\begin{equation}
\hat{N}=\sum_{i=1}^{3}\hat{S}_{ii}
\end{equation}%
commutes with all other operators in the \u3 algebra. This operator is
proportional to the unit operator when acting on occupational states of the
form $|n_{1}n_{2}n_{3}\rangle $,
\begin{equation}
\hat{N}\,|n_{1}n_{2}n_{3}\rangle =A|n_{1}n_{2}n_{3}\rangle \,.
\end{equation}%
Removing $\hat{N}$ reduces \u3 to $\mathfrak{su}(3)$. Thus, the possible
evolutions generated by the Hamiltonian $\hat{H}$ are, up to an unimportant
global phase, finite $\mathfrak{SU}(3 )$ transformations.

\section{Non-Degenerate Case}

\label{nondegenerate}

The non--degeneracy condition is understood to imply that the atomic
transition frequencies $(E_{3}-E_{2})$ and $(E_{2}-E_{1})$ are sufficiently
distinct to satisfy
\begin{equation}
(E_{3}-E_{2})-(E_{2}-E_{1})\gg g_1\, ,\qquad (E_{3}-E_{2})-(E_{2}-E_{1})\gg
g_2\, .
\end{equation}
In this manner, each transition can be interrogated separately by an
external field. A typical non--degenerate system is illustrated in Fig.\ref%
{1atom_nondegenerate}.

\begin{figure}[h]
\begin{center}
\epsfig{file=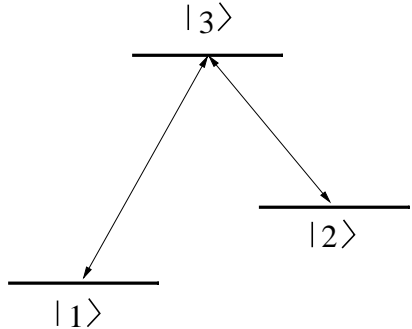, height=1.75in}
\end{center}
\caption{A typical level scheme for the non--degenerate case.}
\label{1atom_nondegenerate}
\end{figure}

For tomographic purposes, the level pairs $(1\leftrightarrow 2)$, $%
(2\leftrightarrow 3)$ and their respective transitions, if taken in
respective isolation, could be considered as independent two--level
subsystems. However the full system must be treated as a three--level system
under $\mathfrak{SU}(3 )$ evolution.

In the rotating frame, the Hamiltonian of Eq.(\ref{hamiltoniangen}) takes
the form
\begin{equation}
\hat{H}_{int}=\Delta _{23}\hat{S}_{33}-\Delta _{12}\hat{S}_{11}+g_{12}\left(
\hat{S}_{12}+\hat{S}_{21}\right) \,+g_{23}\left( \hat{S}_{23}+\hat{S}%
_{32}\right) ,
\end{equation}%
where
\begin{equation}
\Delta _{12}\equiv E_{2}-E_{1}-\omega _{1}\,,\qquad \Delta _{23}\equiv
E_{3}-E_{2}-\omega _{2}\, .
\end{equation}

Because the frequencies $\omega _{1}$ and $\omega _{2}$ of the external
field are adjustable parameters, two types of field pulses can be applied to
our system. The first type is characterized by $\Delta _{ij}\approx 0$ and
thus resonant; it stimulates the corresponding atomic transitions. The
second type is characterized by $\Delta _{ij}\gg g_{ij}$ and is thus
far--off resonant (dispersive); this kind of pulse leads to some phase
shifts.

The corresponding resonant and dispersive evolution operators have the form%
\begin{eqnarray}
U_{ij}^{R}\left( \beta _{ij}\right) &=&\exp\left[-i\beta _{ij}\left( \hat{S}%
_{ij}+\hat{S}_{ji}\right)\right]\, ,\qquad i\ne j\, ,  \label{resonant} \\
U_{11}^{D}\left( \phi _{12}\right) &=&\exp\left[i\phi _{12}\hat{S}_{11}%
\right]\, ,  \label{RD1} \\
U_{33}^{D}\left( \phi _{23}\right) &=&\exp\left[i\phi _{23}\hat{S}_{33}%
\right]\, ,  \label{RD3}
\end{eqnarray}
where $\beta _{ij}=g_{ij}t_{ij}$ and $\phi_{ij}=(\Delta_{ij}+g_{ij}^{2}/%
\Delta _{ij})\tau _{ij}$. Here, $t_{ij},\tau _{ij}$ are time intervals, not
necessarily equal. It should be noted, that for short interaction times $%
\tau_{ij}$ satisfying $g^2\,\tau_{ij}/\Delta_{ij}<<1$ and
$g^2\,\tau_{ij}<< \Delta_{ij}^2$, the second term in the expression
for $\phi _{ij}$ can be obviously neglected. However, for long
interaction times, $\exp (-i\Delta _{ij}\tau _{ij})$ becomes
strongly oscillating and the measurements should be carried out in
stroboscopic times, $\tau _{ij}=2\pi n/\Delta _{ij}$.

For a complete reconstruction of the density matrix in the absence of
degeneracy, it suffices to measure the probability of detecting zero photons
(\emph{i.e.} zero fluorescence condition) in the irradiated field. This
corresponds to the detection the atoms in the ground state, ${|A00\rangle }$%
. It is worth noting here that, in the non-redundant scheme when only a
finite number of different pulses have to be applied, the measurement of
non-zero photons are required \cite{ross}.

It is possible, by combining operations in
Eqs.(\ref{resonant})--(\ref{RD3}), to obtain a element of the group
\U3 sufficiently general for our purpose. This is best seen by first
observing that an element of $\mathfrak{SU}(3 )$ is parameterized by
eight real numbers and
can be conveniently factorized \cite{Hubert} into a product of $\mathfrak{SU}%
(2 )$ subgroup transformations:
\begin{eqnarray}
&& \bar
U(\alpha_{1},\beta_{1},\gamma_1,\alpha_2,\beta_2,\alpha_3,\beta_3,\gamma_3)
\nonumber \\
&&\qquad =R_{23}(\alpha_{1},\beta_{1},\gamma_1)\cdot
R_{12}(\alpha_{2},\beta_{2},\alpha_2)\cdot
R_{23}(\alpha_3,\beta_3,\gamma_3)\, ,
\end{eqnarray}
The action of such a group element on basis states of an irreducible
representation is given in \cite{Hubert}.

The notation
\begin{equation}
n\equiv (n_1n_2n_3)\, ,\qquad I=I_{23}\equiv \textstyle{\frac{1}{2}}%
(n_2+n_3)\, ,  \label{nshorthand}
\end{equation}
will be a useful shorthand throughout this paper. In particular, we note
that
\begin{eqnarray}
\langle \nu \vert\,R_{23}(\alpha,\beta,\gamma)\vert n \rangle&\equiv
\langle
\nu_1\nu_2\nu_3 \vert\,R_{23}(\alpha,\beta,\gamma)\vert n_1n_2n_3 \rangle &=%
\mathcal{D}^{I_{23}}_{M_{\nu},M_n} (\Omega)\, ,  \nonumber \\
\langle \nu \vert R_{12}(\alpha,\beta,\gamma)\vert n \rangle&\equiv
\langle
\nu_1\nu_2\nu_3 \vert\,R_{12}(\alpha,\beta,\gamma)\vert n_1n_2n_3 \rangle &=%
\mathcal{D}^{I_{12}}_{m_{\nu},m_n} (\Omega)\, ,
\end{eqnarray}
where $\mathcal{D}^{J}_{MM^{\prime}}$ an $\mathfrak{SU}(2)$ Wigner $\mathcal{%
D}$-function and
\begin{equation}
I_{12}=\textstyle{\frac{1}{2}}(n_1+n_2)\, ,\quad M_{n}=\textstyle{\frac{1}{2}%
}(n_2-n_3)\, ,\quad m_{n}=\textstyle{\frac{1}{2}}(n_1-n_2)\, .
\label{I12I23}
\end{equation}
More generally, the definition of states for the irrep $(\lambda,\mu)$ and $%
\mathfrak{SU}(3 )$ elements between those states can be found in \cite%
{Hubert}. Specialized results are collected in \ref{Dfunctiondetails}. In
particular, some formulae in this section implicitly depend on matrix
elements of states in irreps of the type $(A,0), (0,A)$ and $(\sigma,\sigma)$%
. The notation of $\mathfrak{SU}(3 )$ $\mathcal{D}$ functions conforms to
that of \cite{Hubert} and uses the pair $(\lambda,\mu)$ with the labels $n,I$
to unambiguously distinguish the $\mathfrak{SU}(3 )$ $\mathcal{D}$
functions. For occupational states, $n,I$ are given by Eq.(\ref{nshorthand}).

Because the state $|A00\rangle $ is, up to a global phase, unchanged by the
action of dispersive pulses and operations of the form $R_{23}$, it is
enough to consider a sequence of pulses of the form:
\begin{equation}
U(\varphi_{23},\beta _{23},\varphi_{12},\beta_{12})
=U_{33}^{D}(-\varphi_{23})\,U_{23}^{R}(\beta_{23})
\,U_{11}^{D}(\varphi_{12})\,U_{12}^{R}(\beta_{12})\, .  \label{ROTADA}
\end{equation}

In the single atom case, the evolution operator Eq.(\ref{ROTADA}) is a $%
3\times 3$ matrix explicitly given by
\begin{eqnarray}  \label{non-deg_1}
&&U(\varphi_{23},\beta _{23},\varphi_{12},\beta_{12})  \nonumber \\
&&=\left({\renewcommand{\arraystretch}{1.5}
\begin{array}{ccc}
\hbox{\rm e}^{i\varphi_{12}}\cos \left(\beta_{12}\right) & -i\,\hbox{\rm e}%
^{i\varphi_{12}}\sin \left(\beta_{12}\right) & 0 \\
-i\,\cos\left(\beta_{23}\right)\sin\left(\beta_{12}\right) &
\cos\left(\beta_{12}\right)\cos\left(\beta_{23}\right) & -i\,\sin\left(%
\beta_{23}\right) \\
-\hbox{\rm e}^{-i\varphi_{23}}\sin\left(\beta_{12}\right)\sin\left(%
\beta_{23}\right) & -i\hbox{\rm e}^{-i\varphi_{23}}\cos \left(\beta
_{12}\right)\sin\left(\beta_{23}\right) & \hbox{\rm e}^{-i\varphi_{23}}\cos
\left( \beta _{23}\right)%
\end{array}%
}\right) \, .  \nonumber \\
\end{eqnarray}
$U(\varphi_{23},\beta _{23},\varphi_{12},\beta_{12})$ can be more easily
analyzed in the factorized form
\begin{equation}
U(\varphi_{23},\beta _{23},\varphi_{12},\beta_{12}) =\hbox{\rm e}^{-\frac{1}{%
3}i(\varphi_{12}-\varphi_{23})}\, \bar{U}\, ,  \label{eq:u3element}
\end{equation}
where
\begin{eqnarray}
\bar{U}&=& \left(%
\begin{array}{ccc}
1 & 0 & 0 \\
0 & -i\hbox{\rm e}^{i(\chi-\varphi_{12})}\cos(\beta_{23}) & \hbox{\rm e}%
^{i(2\chi-\varphi_{12})}\sin(\beta_{23}) \\
0 & -\hbox{\rm e}^{-i(2\chi-\varphi_{12})}\sin(\beta_{23}) & i\hbox{\rm e}%
^{-i(\chi-\varphi_{12})}\cos(\beta_{23})%
\end{array}%
\right)  \nonumber \\
&&\quad \times \left(%
\begin{array}{ccc}
\hbox{\rm e}^{i\chi}\cos(\beta_{12}) & -\sin(\beta_{12}) & 0 \\
\sin(\beta_{12}) & \hbox{\rm e}^{-i\chi}\cos(\beta_{12}) & 0 \\
0 & 0 & 1%
\end{array}%
\right) \left(%
\begin{array}{ccc}
1 & 0 & 0 \\
0 & i\hbox{\rm e}^{i\chi} & 0 \\
0 & 0 & -i\hbox{\rm e}^{-i\chi}%
\end{array}%
\right)\, ,  \label{eq:su3element}
\end{eqnarray}
and
\begin{equation}
\chi=\frac{1}{3}(2\varphi_{12}+\varphi_{23})\, .
\end{equation}
Clearly, the matrix
$U(\varphi_{23},\beta_{23},\varphi_{12},\beta_{12})$ of
Eq.(\ref{eq:u3element}) is an element of \U3 whereas $\bar U$ of
Eq.(\ref {eq:su3element}) is an \Su{3} transformation. Comparing
with the parametrization of
\cite{Hubert}, we have the correspondences
\begin{equation}
\begin{array}{lll}
\alpha_1 \to -\varphi_{23}-\textstyle{\frac{1}{2}}\pi\quad , & \beta_1\to
2\beta_{23}\quad , & \gamma_1\to \frac{3}{2}\pi+\frac{2}{3}\varphi_{12}+%
\frac{1}{3}\varphi_{23}\, \\
\alpha_2\to-\frac{2}{3}\varphi_{12}-\frac{1}{3}\varphi_{23}\, , & \beta_2\to
2\beta_{12}\, ,\quad &  \\
\alpha_3\to-\frac{4}{3}\varphi_{12}-\frac{2}{3}\varphi_{23}-\pi\, ,\quad &
\beta_3=0\, , & \gamma_3=0\, .%
\end{array}%
\end{equation}
Note that, although $\bar{U}$ is not the most general \Su{3}, it can
be multiplied on the right by a transformation of the form
\begin{eqnarray}
&&R_{12}(\bar\alpha_2,0,\bar\alpha_2)R_{23}(\bar\alpha_3,\bar\beta_3,\bar%
\gamma_3)=  \nonumber \\
&&\qquad \left(%
\begin{array}{ccc}
\hbox{\rm e}^{-i\bar\alpha_2} & 0 & 0 \\
0 & \hbox{\rm e}^{\frac{i}{2}i(2\bar\alpha_2-\bar\alpha_3-\bar\gamma_3)}\cos(%
\frac{\bar\beta_3}{2}) & -\hbox{\rm e}^{\frac{i}{2}(2\bar\alpha_2-\bar%
\alpha_3+\bar\gamma_3)}\sin(\frac{\bar\beta_3}{2}) \\
0 & \hbox{\rm e}^{\frac{i}{2}(\bar\alpha_3-\bar\gamma_3)}\sin(\frac{%
\bar\beta_3}{2}) & \hbox{\rm e}^{\frac{i}{2} (\bar\alpha_3+\bar\gamma_3)}%
\cos(\frac{\bar\beta_3}{2})%
\end{array}%
\right)
\end{eqnarray}
without affecting the dynamics of the $\vert 100 \rangle$ state. Thus, $\bar{%
U}$ is equivalent to a general transformation when acting on $\vert 100
\rangle$.

Expanding the density matrix in the occupational basis:
\begin{equation}
\rho =\sum_{n\nu}\,\vert n \rangle\langle \nu \vert\,\rho_{n,\nu}\, ,
\end{equation}
and introducing the shorthand $\tau =(\varphi_{23},\beta
_{23},\varphi_{12},\beta_{12})\, ,$ we rapidly obtain
\begin{eqnarray}
\omega (\tau ) &=&\sum_{n\nu }\,\langle A00|\,\bar{U}(\tau )\,|n\rangle
\langle \nu |\bar{U}^{\dagger }(\tau )\,|A00\rangle \,\rho_{n,\nu}\,
\nonumber \\
&=&\sum_{n\nu }\,(-1)^{\nu _{2}}\,\mathcal{D}_{(A00)0,nI}^{(A,0)}(\tau )\,%
\mathcal{D}_{(0AA)0,\nu^*I^\prime}^{(0,A)}(\tau)\rho_{n,\nu},
\label{omegaasproduct}
\end{eqnarray}
where
\begin{equation}
\mathcal{D}^{(\lambda,\mu)}_{n_1\bar{I}_1,n_2I_2}(\tau)\equiv \langle
(\lambda,\mu)n_1 I_1 \vert\,U(\tau)\vert (\lambda,\mu)n_2 I_2 \rangle
\end{equation}
is an $\mathfrak{SU}(3 )$ Wigner-function for the irrep $(\lambda,\mu)$.
Further notational details and properties of these functions (in particular
Eq.(\ref{Dfunctionstar})) can be found in \ref{Dfunctiondetails}.

Products of $\mathfrak{SU}(2 )$ $\mathcal{D}$-functions can be decomposed
into sums of $\mathcal{D}$-functions multiplied by products of $\mathfrak{SU}%
(2 )$ Clebsch-Gordan coefficients. The same holds for products of $\mathfrak{%
SU}(3 )$ $\mathcal{D}$-functions provided that we use $\mathfrak{SU}(3 )$
Clebsch-Gordan technology. Thus, given that $\mathfrak{SU}(3)$-coupling $%
(A,0)\otimes(0,A)$ decomposes in the direct sum \cite{su3tensor}
\begin{eqnarray}
(A,0)\otimes(0,A)&=&(A,A)\oplus(A-1,A-1)\oplus\ldots\oplus (0,0)\, ,
\nonumber \\
&=&\bigoplus_{\lambda=0}^{A}\,(\lambda,\lambda)\, ,
\end{eqnarray}
we have
\begin{eqnarray}
&&\mathcal{D}_{(A00)0,\,nI}^{(A,0)}(\tau )\, \mathcal{D}_{(0AA)0,\,\nu^*I^%
\prime}^{(0,A)}(\tau )  \nonumber \\
&&=\sum_{\lambda,J} \mathcal{D}^{(\lambda,\lambda)}_{(\lambda\lambda%
\lambda)0,\,N^{(\lambda)} J}(\tau)\,
\sucg{(A,0)}{(A00)0}{(0,A)}{(0AA)0}{(\lambda,\lambda)}{(\lambda\lambda\lambda)0}
\nonumber \\
&&\qquad\qquad\times
\sucg{(A,0)}{nI}{(0,A)}{\nu^*I^{\prime}}{(\lambda,\lambda)}{N^{(\lambda)}J}\,
,  \label{su3Dcombined}
\end{eqnarray}
where
\begin{equation}
N^{(\lambda)}=(n_1+\nu_1^*-(A-\lambda),n_2+\nu_2^*-(A-\lambda),n_3+%
\nu_3^*-(A-\lambda))\, .  \label{Nlambdadef}
\end{equation}
and where
\begin{equation}
\sucg{(A,0) }{n_1I_1}{(0,A) }{n_2
I_2}{(\lambda,\lambda)}{N^{(\lambda)} I_3}
\end{equation}
is the $\mathfrak{SU}(3 )$ Clebsch-Gordan coefficient for the coupling of $%
\vert (A,0)\,n_1I_1 \rangle$ and $\vert (0,A)\,n_2I_2 \rangle$ to $\vert
(\lambda,\lambda)\,N^{(\lambda)} I_3 \rangle$. The appearance of extra
factors of $(A-\lambda)$ in the construction of $N^{(\lambda)}$ is discussed
in Eq.(\ref{Nsigmadef}) of \ref{su3CGcoeffs}.

Inserting Eq.(\ref{su3Dcombined}) in Eq.(\ref{omegaasproduct}) yields
\begin{eqnarray}
\omega(\tau) &=&\displaystyle\sum_{n\nu\lambda J}
\,(-1)^{\nu_2}\,\rho_{n,\nu}\, \mathcal{D}^{(\lambda,\lambda)}_{(\lambda%
\lambda\lambda) 0,\,N^{(\lambda)} J}(\tau)  \nonumber \\
&&\ \times
\sucg{(A,0)}{(A00)0}{(0,A)}{(0AA)0}{(\lambda,\lambda)}{(\lambda\lambda\lambda)0}
\,
\sucg{(A,0)}{nI}{(0,A)}{\nu^*I^\prime}{(\lambda,\lambda)}{N^{(\lambda)}J}
\, .  \label{eqnondgtomo}
\end{eqnarray}
After some straightforward manipulations detailed in
\ref{sec:casenondegenerate}, we obtain the final expression
\begin{eqnarray}
&&(-1)^{\nu_2}\rho_{n,\,\nu}=\sum_{\mu\,J}\frac{(\mu+1)^3}{1024\pi^5}\,
\sucg{(A,0)}{(A00)0}{(0,A)}{(0AA)0}{(\mu,\mu)}{(\mu\mu\mu)\,0}^{-1}\,  \nonumber \\
&&\qquad\qquad \times
\sucg{(A,0)}{nI}{(0,A)}{\nu^*I^{\prime}}{(\mu,\mu)}{N^{(\mu)} J}
\, \int d\Omega\, \mathcal{D}^{(\mu,\mu)*}_{(\mu\mu\mu)\,0,%
\,N^{(\mu)} J}(\tau)\,\omega(\tau)\, .  \label{reconsfornondeg}
\end{eqnarray}

As there is no restriction on $n$ or $\nu^*$, Eq.(\ref{reconsfornondeg})
shows that, in the non-degenerate case, the density matrix can be completely
reconstructed.

\section{Degenerate $\Lambda $-type atomic systems}

\label{sec:lambdacase}

Let us turn our attention to the case of a degenerate $\Lambda$-type system.
A typical $\Lambda$ atom is schematically illustrated in Fig.\ref%
{Lambdafigure}.

\begin{figure}[h]
\begin{center}
\epsfig{file=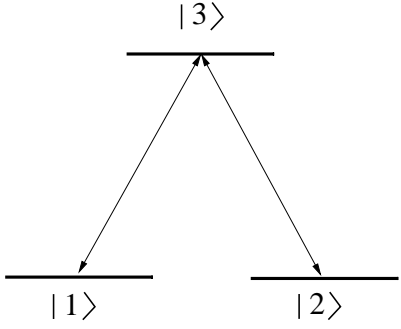, height=1.75in}
\end{center}
\caption{Schematic representation of a single atom in a $\Lambda$
configuration.}
\label{Lambdafigure}
\end{figure}

In the single--atom case, the allowed transitions are $\left\vert
1\right\rangle \leftrightarrow \left\vert 3\right\rangle ,\left\vert
2\right\rangle \leftrightarrow \left\vert 3\right\rangle .$ The degeneracy
condition is
\begin{equation}
E_{3}-E_{1}=E_{3}-E_{2}.
\end{equation}

In the multi--atom case, the only atomic configuration that can be
unambiguously identified by photon counting is when every atom is excited,
\emph{i.e.} when the system of $A$ atoms is in the state $|00A\rangle $.

\subsection{The evolution}

In the rotating frame, the interaction Hamiltonian has the form
\begin{equation}
\hat{H}_{\Lambda }=\Delta S_{33}+g\left( S_{13}+S_{23}\right) +g\left(
S_{31}+S_{32}\right) ,  \label{hamiltonianlambda}
\end{equation}%
where $\Delta =E_{3}-E_{1}-\omega $ and $g_{1}=g_{2}=g$ for simplicity.

In the single atom case, $\hat{H}_{\Lambda}$ can be represented as the
following $3\times 3$ matrix:
\begin{equation}
\hat{H}_{\Lambda }=\left( {\renewcommand\arraycolsep{0.5em}%
\begin{array}{ccc}
0 & 0 & g \\
0 & 0 & g \\
g & g & \Delta%
\end{array}%
}\right) \,.  \label{H_L1}
\end{equation}

A simple basis transformation
\begin{eqnarray}
|\,1\rangle &\mapsto &\vert \,\tilde{1} \rangle =\frac{1}{\sqrt{2}}\,
\left(\vert \,1 \rangle - \vert \,2 \rangle\right)\,,\quad \vert \,2 \rangle
\mapsto \vert \,\tilde{2} \rangle =\frac{1}{\sqrt{2}}\,\left(\vert \,1
\rangle + \vert \,2 \rangle\right) \, ,  \nonumber \\
\vert \,3 \rangle &\mapsto& \vert \,\tilde{3} \rangle \,,
\label{tildestates}
\end{eqnarray}
given by the constant matrix
\begin{equation}
T_{12}=\left( {\renewcommand\arraycolsep{0.4em}%
\begin{array}{ccc}
\textstyle\frac{1}{\sqrt{2}} & \textstyle\frac{1}{\sqrt{2}} & 0 \\
-\textstyle\frac{1}{\sqrt{2}} & \textstyle\frac{1}{\sqrt{2}} & 0 \\
0 & 0 & 1%
\end{array}%
}\right) ,
\end{equation}%
transforms Eq.(\ref{H_L1}) to the block diagonal form
\begin{equation}
\hat{H}_{\Lambda }\mapsto \hat{H}_{T}=T_{12}^{-1}\,\hat{H}_{\Lambda
}\,T_{12}=\left( {\renewcommand\arraycolsep{0.4em}%
\begin{array}{ccc}
0 & 0 & 0 \\
0 & 0 & \sqrt{2}g \\
0 & \sqrt{2}g & \Delta%
\end{array}%
}\right) .
\end{equation}

The effect $T_{12}$ on basis states is illustrated in Fig.\ref%
{Lambda_atom_dark}; $T_{12}$ produces a \emph{dark state} $\vert \tilde 1
\rangle$ completely decoupled from the remaining doublet. In view of this we
can expect, on general grounds, that a complete reconstruction will not be
possible as our Hamiltonian $\hat{H}_{\Lambda }$ cannot possibly probe the
dark state $|\tilde{1}\rangle $.

\begin{figure}[h]
\begin{center}
\epsfig{file=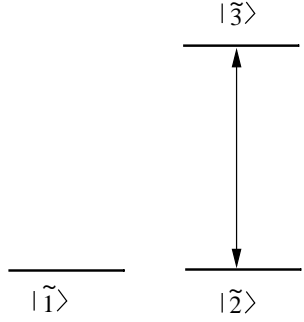, height=1.75in}
\end{center}
\caption{The basis states resulting from the transformation $T_{12}$.}
\label{Lambda_atom_dark}
\end{figure}

Using the basis $\{\vert \tilde 1 \rangle,\vert \tilde 2 \rangle,\vert
\tilde 3 \rangle\}$, the resonant pulses, with $\Delta=0$, are of the form
\begin{equation}
\tilde U_{\Lambda}^R(\sqrt{2}gt)=\left(%
\begin{array}{ccc}
1 & 0 & 0 \\
0 & \cos\left(\sqrt{2}gt\right) & i\sin\left(\sqrt{2}gt\right) \\
0 & i\sin\left(\sqrt{2}gt\right) & \cos\left(\sqrt{2}gt\right)%
\end{array}%
\right)\, .  \label{lambdaresonant}
\end{equation}

In the same basis, the dispersive pulses, with $\Delta \gg g$, are described
in the stroboscopic approximation by the effective evolution operator%
\begin{equation}
\tilde{U}_{\Lambda }^{D}\left( 2g^{2}t/\Delta \right) =\left(
\begin{array}{ccc}
1 & 0 & 0 \\
0 & \hbox{\rm e}^{2ig^{2}t/\Delta } & 0 \\
0 & 0 & \hbox{\rm e}^{-2ig^{2}t/\Delta }%
\end{array}
\right) .  \label{lambdadispersive}
\end{equation}

In the two--dimensional subspace spanned by $|\tilde{2}\rangle $ and $|%
\tilde{3}\rangle $, the operators $\tilde{U}_{\Lambda }^{R}$ and $\tilde{U}%
_{\Lambda }^{D}$ correspond to $\mathfrak{SU}(2)$ rotations about
the
$\hat{x}$ and
$\hat{z}$ axes, respectively:
\begin{equation}
\tilde{U}_{\Lambda }^{R}(\alpha )\mapsto R_{23}^{x}(\alpha )\,,\qquad \tilde{%
U}_{\Lambda }^{D}(\beta )\mapsto R_{23}^{z}(\beta )\,,
\end{equation}%
in an obvious notation. A sufficiently general sequence of pulses can thus
be written
\begin{eqnarray}
\tilde{U}_{\Lambda }(\tilde{\Omega}) &=&R_{23}^{z}(\alpha )\cdot
R_{23}^{x}(\beta )\cdot R_{23}^{z}(\gamma )\,, \nonumber \\
&=&R_{23}^{z}(\alpha +\textstyle\frac{\pi }{2})\cdot R_{23}^{y}(\beta )\cdot
R_{23}^{z}(\gamma -\textstyle\frac{\pi
}{2})\,=R_{23}(\tilde{\Omega})\,. \label{lambdapulse}
\end{eqnarray}%
For the one-atom case, the $3\times 3$ matrix representation for this
evolution has the form
\begin{equation}
\tilde{U}_{\Lambda }(\tilde{\Omega})=\left( {\renewcommand{%
\arraycolsep}{0.75em}%
\begin{array}{ccc}
1 & 0 & 0 \\
0 & \ast & \ast \\
0 & \ast & \ast%
\end{array}%
}\right)
\end{equation}%
where $\ast $ indicates a non-zero entry. The block diagonal form of $\tilde{%
U}$ is explicit. It shows that, for a system containing one or more
than one atom, there will always be at least one subspace which
cannot be reached in the course of evolution of $\vert 00A \rangle$;
such decoupled subspaces are an obstruction to complete
reconstruction.

To illustrate this, we expand the density matrix for an $A$-atom
system in the basis
$\{|\tilde{n}_{1}\tilde{n}_{2}\tilde{n}_{3}\rangle \}$ of occupation
of the states $|\tilde{1}\rangle \,,|\tilde{2}\rangle $ and
$|\tilde{3}\rangle $:
\begin{equation}
\tilde{\rho}=\sum_{\tilde{n}_{1}\tilde{n}_{2}\tilde{n}_{3}\tilde{\nu}_{1}%
\tilde{\nu}_{2}\tilde{\nu}_{3}}|\tilde{n}_{1}\tilde{n}_{2}\tilde{n}%
_{3}\rangle \langle \tilde{\nu}_{1}\tilde{\nu}_{2}\tilde{\nu}_{3}|\,\tilde{%
\rho}_{(\tilde{n}_{1}\tilde{n}_{2}\tilde{n}_{3}),(\tilde{\nu}_{1}\tilde{\nu}%
_{2}\tilde{\nu}_{3})}\, .  \label{rho_l}
\end{equation}%
Using the shorthand $\tilde{n}$ for the triplet $\tilde{n}_{1}\tilde{n}%
_{2}\tilde{n}_{3}$, we can write the tomogram as
\begin{equation}
\omega(\tilde{\Omega})=\sum_{\tilde{n}\tilde{\nu}}\langle 00A\vert R_{23}(%
\tilde{\Omega})\,|\tilde{n}_{1}\tilde{n}_{2}\tilde{n}_{3}\rangle \langle
\tilde{\nu}_{1}\tilde{\nu}_{2}\tilde{\nu}_{3}\vert R_{23}(\tilde{\Omega}%
)|00A\rangle \,\tilde{\rho}_{(\tilde{n}_{1}\tilde{n}_{2}\tilde{n}_{3}),(%
\tilde{\nu}_{1}\tilde{\nu}_{2}\tilde{\nu}_{3})}\,.  \label{tomolambdasystem}
\end{equation}

As the $R_{23}(\bar{\Omega})$ rotation does not affect the first atomic
index and acts irreducibly as an $\mathfrak{SU}(2 )$ rotation in each the
subspace spanned by states having a common fixed $\tilde n_1$, we write
\begin{equation}
\vert \tilde n_1\tilde n_2\tilde n_3 \rangle\to \vert \tilde{n}_1; Im
\rangle\, ,\qquad \vert \,\tilde{\nu}_1\tilde{\nu}_2\tilde{\nu}_3 \rangle\to
\vert \,\tilde \nu_1 ;I\mu \rangle\, ,  \label{Lambda_su2notation}
\end{equation}
where
\begin{equation}
I=\textstyle{\frac{1}{2}}(\tilde{n}_{2}+\tilde{n}_{3}) =\textstyle{\frac{1}{2%
}}(\tilde{\nu}_{2}+\tilde{\nu}_{3})\,, \quad m=\textstyle{\frac{1}{2}}(%
\tilde{n}_{2}-\tilde{n}_{3})\,,\quad \mu =\textstyle{\frac{1}{2}}(\tilde{\nu}%
_{2}-\tilde{\nu}_{3})\,  \label{Ilambdasystem}
\end{equation}

Furthermore, the tomograms of Eq.(\ref{tomolambdasystem}) must have $\tilde{n%
}_{1}=\tilde{\nu}_{1}=0$, so Eq.(\ref{Ilambdasystem}) leads to
\begin{eqnarray}
\omega (\tilde{\Omega}) &=&\sum_{m\,\mu } \mathcal{D}_{-I,\,m}^{I}(\tilde{%
\Omega})\, \mathcal{D}_{-I,\,\mu}^{I\ast}(\tilde{\Omega})\, \tilde{\rho}%
_{(0;Im),(0;I\mu)}  \nonumber \\
&=&\sum_{m\,\mu L}\,(-1)^{-I-\mu }\,C^{L\,0}_{I\,I,\, I\,-I}
C^{L\,M}_{I\,m,\, I\,-\mu } \,\mathcal{D}_{0\,,M}^{L}(\tilde{\Omega})\,\,%
\tilde{\rho}_{(0;Im),(0;I\mu )}\,,  \label{tom_l_3}
\end{eqnarray}%
where $\tilde{\rho}_{(n_{1}n_{2}n_{3}),(\nu _{1}\nu _{2}\nu _{3})}\to \tilde{%
\rho}_{(n_{1};Im),(\nu _{1};I\mu )}$ has been used to conform to the
notation of Eq.(\ref{Lambda_su2notation}), where $\mathcal{D}^I_{mm^{\prime}}
$ is the usual $\mathfrak{SU}(2 )$ $\mathcal{D}$ function and where $%
C^{L\,M}_{L_{1}\,m_{1},\, L_{2}\,m_{2}}$ is an $\mathfrak{SU}(2 )$
Clebsch-Gordan coefficient. 

Multiplying both sides of (\ref{tom_l_3}) by $\mathcal{D}_{0,M}^{L^{\prime
}\ast }(\tilde{\Omega})$, integrating over $\mathfrak{SU}(2 )$ and using
orthogonality of the Clebsch-Gordan coefficients rapidly gives 
the elements of the density matrix that can be reconstructed from the
tomographic process as
\begin{eqnarray}
\,\tilde{\rho}_{(0;Im),(0;I\mu )}&=&(-1)^{I+\mu }\sum_{L}\frac{2L+1}{8\pi
^{2}}\,C^{L\,M}_{I\,m,\, I\,-\mu } \left(C^{L\,0}_{I\,I,\, I\,-I}\right)^{-1}
\nonumber \\
&&\qquad\qquad \times \int \,d\tilde{\Omega}\,\omega (\tilde{\Omega})\,
\mathcal{D}_{0,M}^{L\ast}(\tilde{\Omega}).
\end{eqnarray}

\subsection{Reconstruction for state of one and two $\Lambda $-type atoms.}

In a system containing a single atom, the Hilbert space is spanned, in the
notation of Eq.(\ref{Lambda_su2notation}), by states of the form $\vert
\tilde n_1; Im \rangle$, with
\begin{equation}
\vert \tilde 1 \rangle=\vert 100 \rangle\, ,\qquad \vert \tilde 2
\rangle=\vert 0;\textstyle{\frac{1}{2}}\textstyle{\frac{1}{2}} \rangle\,
,\qquad \vert \tilde 3 \rangle=\vert 0;\textstyle{\frac{1}{2}},-\textstyle{%
\frac{1}{2}} \rangle\, .
\end{equation}
Under the evolution $\tilde U_{\Lambda}(\tilde \Omega)=R_{23}(\tilde\Omega)$%
, the initial state $\vert 0;\textstyle{\frac{1}{2}},-\textstyle{\frac{1}{2}}
\rangle$ cannot reach the dark state so it is only possible to reconstruct
element of $\rho$ of the form $\rho_{(0;\frac{1}{2} m),(0;\frac{1}{2}
m^{\prime})}$, with $m,m^{\prime}=\pm\,\textstyle{\frac{1}{2}}$. The last
diagonal element, $\rho_{(1;00),(1;00)}$ can be inferred from the
normalization. None of the remaining four matrix elements can be determined
by our scheme.

In the two-atom case, an even smaller proportion of matrix elements can be
recovered. Using again the notation of Eq.(\ref{Lambda_su2notation}), states
of the irrep $(2,0)$ are conveniently given, in the occupational, tensor
product and $\mathfrak{SU}(2 )$ basis $\vert \tilde n_1; \ell m \rangle$, in
table \ref{twoatomlambda}.

\begin{table}[htp]
\caption{Basis states for two atoms in the $\Lambda$ configuration.}
\label{twoatomlambda}
\begin{center}
\begin{tabular}{|c||c||c|}
\hline
$\quad \vert n_1n_2n_3 \rangle$\quad & $\qquad\vert \tilde 1
\rangle^{n_1}\vert 2 \rangle^{n_2}\vert 3 \rangle^{n_3}\qquad $ & $%
\quad\vert \tilde n_1;I,m \rangle\quad $ \\ \hline\hline
$\vert 002 \rangle $ & $\vert \tilde 3 \rangle\vert \tilde 3 \rangle $ & $%
\vert \,0;1,-1 \rangle $ \\
$\vert 011 \rangle $ & $\frac{1}{\sqrt{2}}\left(\vert \tilde 2 \rangle\vert
\tilde 3 \rangle+\vert \tilde 3 \rangle\vert \tilde 2 \rangle\right)$ & $%
\vert \,0;1,0 \rangle $ \\
$\vert 020 \rangle $ & $\vert \tilde 2 \rangle\vert \tilde 2 \rangle $ & $%
\vert \,0;1,1 \rangle $ \\
$\vert 101 \rangle $ & $\frac{1}{\sqrt{2}}\left(\vert \tilde 1 \rangle\vert
\tilde 3 \rangle+\vert \tilde 3 \rangle\vert \tilde 1 \rangle\right)$ & $%
\vert\, 1;\textstyle{\frac{1}{2}},-\textstyle{\frac{1}{2}} \rangle$ \\
$\vert 110 \rangle $ & $\frac{1}{\sqrt{2}}\left(\vert \tilde 1 \rangle\vert
\tilde 2 \rangle+\vert \tilde 2 \rangle\vert \tilde 1 \rangle\right) $ & $%
\vert\, 1;\textstyle{\frac{1}{2}},\textstyle{\frac{1}{2}} \rangle $ \\
$\vert 200 \rangle$ & $\vert \tilde 1 \rangle\vert \tilde 1 \rangle$ & $%
\vert \,2;0,0 \rangle$ \\ \hline
\end{tabular}%
\end{center}
\end{table}

The initial state $\vert 002 \rangle$ will not evolve out of the
$I=1$ subspace, so only matrix elements of the form
$\rho_{(0;1m)(0;1m^{\prime})}$ can be reconstructed using our
scheme. These represent only nine of the possible 36 elements of the
density matrix.

We conclude this section by noting that the situation obviously worsens (in
the sense that a smaller and smaller proportions of the matrix elements can
be recovered) as the number of atoms increases.

\section{Degenerate $\Xi $-type atomic systems}

\label{sec:xicase}

\begin{figure}[h]
\begin{center}
\epsfig{file=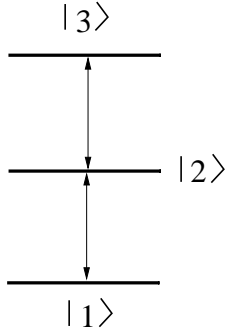, height=1.75in}
\end{center}
\caption{The $\Xi$ configuration for a single atom.}
\label{1atom_Sigma}
\end{figure}

Finally, we consider the case of the $\Xi$ system. It is illustrated, for a
single atom, in Fig.\ref{1atom_Sigma}. For this configuration, the condition
$E_{2}-E_{1}=E_{3}-E_{2}$ holds.

\subsection{The evolution}

In the rotating frame, the Hamiltonian governing the evolution of a
collection of $A$ atoms in the $\Xi$ configuration in an external field has
the form $\left( g_{1}=g_{2}=g\right) $
\begin{equation}
\hat{H}_{\Xi }=\Delta \left( \hat{S}_{33}-\hat{S}_{11}\right) +g\left( \hat{S%
}_{12}+\hat{S}_{32}+\hat{S}_{21}+\hat{S}_{23}\right) ,  \label{Hsig}
\end{equation}%
where $\Delta =(E_{3}-E_{1})/2-\omega $.

Important insight into the nature of this Hamiltonian can be gained by
noting that the operators $\hat{S}_{11}-\hat{S}_{33}$ and $\hat{S}_{12}+\hat{%
S}_{21}+\hat{S}_{23}+\hat{S}_{32}$ are, in fact, proportional to two of the
three generators of the \so3 subalgebra of $\mathfrak{su}(3 )$ :
\begin{equation}
\hat{S}_{11}-\hat{S}_{33}\mapsto \hat{L}_{z}\,,\qquad \hat{S}_{12}+\hat{S}%
_{21}+\hat{S}_{23}+\hat{S}_{32}\mapsto \sqrt{2}\,\hat{L}_{x}\,.
\end{equation}
Thus, the possible evolutions are elements of the $\mathfrak{SO}(3)$
subgroup of $\mathfrak{SU}(3)$.

Clearly, a convenient sequence of pulse is given by
\beqa
U_{\Xi}(\Omega)&\equiv &R^z(\alpha)\cdot R^x(\beta)\cdot
R^z(\gamma)\, ,\nonumber
\\
&=&R^z(\alpha+\pi/2)\cdot R^y(\beta)\cdot R^z(\gamma-\pi/2)\,
.\label{Sigmapulsesequence}
\eeqa
Here, the resonant pulses are of the form
\beq
R^x(\beta)=\exp\left(-i\beta\,(\hat{S}_{12}+\hat{S}%
_{21}+\hat{S}_{23}+\hat{S}_{32})/\sqrt{2}\right)
\eeq
while the dispersive pulses are generated by $\hat S_{11}-\hat
S_{33}$.

It is important to note that, in the one--atom case, the
exponentiation of
$\hat H_{\Xi}$ in Eq.(\ref{Hsig}) produces an evolution that acts
irreducibly on the non--degenerate states of the Hilbert space: in
contrast with Eq.(\ref{lambdapulse}) of the $\Lambda$ case, the
``rotations''
$R^x$ and $R^z$ of the $\Xi$ states are not restricted to a
two--dimensional subspace of the whole Hilbert space.

To analyze the many-atom case, we start by observing that the state
of the system for which every the atom is completely excited,
$\left\vert 00A\right\rangle $, is an eigenstate of $\hat L_{z}$
with eigenvalue $-A$ and is annihilated by $\hat L_-$. Here,
$\hat L_-=\left(\hat S_{21}+\hat S_{32}\right)/\sqrt{2}$ is constructed in
the usual way: $\hat L_-=\hat L_x-i\hat L_y$. Thus, $\left\vert
00A\right\rangle$ is the unique angular momentum state $\left\vert
L,-L\right\rangle $, with $L=A$:
\begin{equation}
\vert 00A \rangle\to \vert L=A,M=A \rangle\, .
\end{equation}
As this state contains the largest possible number of excitations, it can be
uniquely identified through photon counting so that the corresponding
tomogram is determined from the probability of detecting $2A$ photons in the
irradiated field.

The general correspondence between the occupational basis states $\left\vert
n_{1}n_{2}n_{3}\right\rangle $ is found in \cite{SharpNPA129} and given by
\begin{eqnarray}
&&\vert LM \rangle=\sqrt{\frac{2^{L+M}\left(\textstyle{\frac{1}{2}}%
(A+L)\right)!(L+M)!(L-M)!(2L+1)}{\left(\textstyle{\frac{1}{2}}%
(A-L)\right)!(A+L+1)!}}  \nonumber \\
&&\qquad\quad\times
\left((a_2^{\dagger})^2-2a_{1}^{\dagger}a_3^{\dagger}\right)^{\frac{1}{2}%
(A-L)} \displaystyle\sum_p \frac{(a_1^{\dagger})^p\,(a_2^{\dagger})^{L+M-2p}%
\,(a_3^{\dagger})^{p-M}}{2^p \, p! \,(p-M)!(L+M-p)!}\,\vert 0 \rangle\, .
\end{eqnarray}

\begin{table}[htp]
\caption{Angular momentum basis states as linear combinations of
occupational number states for one and two atoms in the $\Xi$ configuration.}
\label{so3insu3states}
\begin{center}
\begin{tabular}{|c||c|c||c|}
\hline
$\quad A\quad $ & $\quad L\quad $ & $\quad M \quad $ & $\quad \vert LM
\rangle\quad $ \\ \hline\hline
1 & 1 & $1$ & $a_{1}^\dagger \vert 0 \rangle$ \\ \hline
&  & 0 & $a_2^\dagger\vert 0 \rangle$ \\ \hline
&  & $-1$ & $a_{3}^\dagger \vert 0 \rangle$ \\ \hline\hline
2 & 2 & $2$ & $\frac{1}{\sqrt{2}}{(a_{1}^\dagger)^2}\, \vert 0 \rangle$ \\
\hline
&  & $1$ & $a_2^{\dagger}a_{ 1}^{\dagger}\vert 0 \rangle$ \\ \hline
&  & $0$ & $\frac{1}{\sqrt{3}}\left((a_2^\dagger)^2+a_1^\dagger
a_3^\dagger\right)\vert 0 \rangle$ \\ \hline
&  & $-1$ & $a_2^{\dagger}a_{3}^{\dagger}\vert 0 \rangle$ \\ \hline
&  & $-2$ & $\frac{1}{\sqrt{2}}{(a_{3}^\dagger)^2}\, \vert 0 \rangle$ \\
\hline\hline
2 & 0 & $0$ & $\frac{1}{\sqrt{6}}\left( (a_2^\dagger)^2-2a_1^\dagger
a_3^\dagger\right)\vert 0 \rangle$ \\ \hline
\end{tabular}%
\end{center}
\end{table}

It is clear that, given an angular momentum state in the irrep $(A,0)$ of $%
\mathfrak{su}(3)$, we can unambiguously write it as a linear combination of
occupational states, and vice versa. Thus, we may expand%
\begin{equation}
\rho =\sum_{L_{1}M_{1}L_{2}M_{2}}\vert L_{1}M_{1} \rangle\langle L_{2}M_{2}
\vert\, \rho_{L_{1}M_{1},L_{2}M_{2}},
\end{equation}
where $L_{1},L_{2}$ run from $A,A-2,...,1$ or $0$ depending if $A$ is even
or odd.

As the evolution is necessarily an element of \So3, the tomogram takes the
general form
\begin{eqnarray}
\omega (\Omega ) &=&\sum_{L_1M_1L_2M_2}\langle A,-A \vert R(\Omega)\vert
L_1M_1 \rangle \langle L_2M_2 \vert R^{\dagger}(\Omega )\vert A,-A \rangle\,
\rho_{L_1M_1,L_2M_2},  \nonumber \\
&=&\sum_{M_1M_2}\mathcal{D}_{-A,M_1}^{A}(\Omega )\mathcal{D}_{-A,M_2}^{A\,*
}(\Omega )\,\rho_{AM_1,AM_2},  \nonumber \\
&=&\sum_{M_1M_2J}\,(-1)^{-A-M_2}\,\mathcal{D}_{0,M}^{J}(\Omega)
C^{J\,0}_{A\,-A,\, A\,A} C^{J\,M}_{A\,M_1,\, A\,M_2} \,\rho_{AM_1,AM_2},
\end{eqnarray}%
with $C^{J\,M}_{L_1 \,M_1,\, L_2\,M_2}$ a regular angular momentum
Clebsch-Gordan coefficient.

In a manner similar to the previous cases, multiplication by $\mathcal{D}%
_{0,M}^{J^{\prime }\ \ast }(\Omega )$, integration over $\mathfrak{SO}(3)$
and orthogonality of Clebsch-Gordan coefficients yields
\begin{eqnarray}
&&(-1)^{A+M_{2}}\,\rho _{AM_{1}AM_{2}}  \nonumber \\
&&\qquad =\sum_{J}\,\frac{2J+1}{8\pi ^{2}}\,C_{A\,M_{1},\,A\,M_{2}}^{J\,M}%
\left( C_{A\,-A,\,A\,A}^{J\,0}\right) ^{-1}\,\int d\Omega \,\mathcal{D}%
_{0,M}^{J\ast }(\Omega )\omega (\Omega )\,. \label{rho_sigma}
\end{eqnarray}%
The result clearly shows that only those linear combinations of occupational
state that transform by angular momentum $L=A$ can be reconstructed.

\subsection{Examples: one and two atom cases}

For a single atom, we see from table \ref{so3insu3states}, that the tomogram
is constructed from an $L=1$ state. There is no other angular momentum
multiplet and so the evolution, an element of \So3, will yield sufficiently
many tomograms to guarantee complete reconstruction.

The matter is different for the two-atom case. In this case, the tomogram is
constructed using an $L=2$ state but the Hilbert space also contains an $L=0$
subspace, which cannot be reached from $L=2$ with our evolution. Thus, if we
are limited to measuring a total of $2A$ photons, it will only be possible
to recover $\rho_{2M,2M^{\prime}}$ and impossible to reconstruct $%
\rho_{00,2M},\rho_{00,00},\rho_{2M,00}$. This is because, in our scheme, it
is not possible to extract photons from the $L=0$ state, and absence of
photon does not pin down a particular state.

It may be possible to measure fewer than $2A$ photons, but this does not
lead to more information. There is only one state with $M=A$ and one state
with $M=A-1$ (or $M=-A$ and $M=-(A-1))$. It is possible to use the $L=A,M=A-1
$ state for the tomogram and measure $2A-1$ or $2A-2$ photons, but we will
recover nothing more than if we had started with $L=A,M=A$.

There are two states with $M=A-2$; they belong different angular momentum
multiplet. Thus, if we measure, say, a total of $2(A-2)$ photons, it is not
possible to know unambiguously if this is the result of a complete cascade
within the $L=A-1$ multiplet or a partial cascade within the $L=A$
multiplet. This kind of limitation becomes obviously more severe as the
number of angular momentum multiplet containing a given $M$ value increases.

\section{Conclusions}

We have proposed a physical realization applicable to the
reconstruction of the quantum state of three-level atomic systems.
The information about atomic states is extracted by measuring the
total number of excitations after successive applications of
electromagnetic field pulses.

We have shown that, in the non-degenerate case, the complete
reconstruction of atomic states is possible. Although the number of
independent parameters required for a complete reconstruction is
less than needed for the complete parametrization of a generic
element of $\mathfrak{SU}(3)$ group, a complete reconstruction is
possible because, in addition to the usual evolution of the system,
another tool is available in the reconstruction scheme: the
projective measurement.

When degeneracies are present, the possibilities of reconstruction
are limited. The origin of these limitations is essentially
different for atoms in $\Lambda $ and $\Xi$ configuration.  In both
cases, the evolution operator operators belong to a subgroup of the
whole $\mathfrak{SU}(3)$ group, and our work illuminates the subtle
distinction between the global properties of $\mathfrak{SO}(3)$ and
$\mathfrak{SU}(2)$ as subgroups of $\mathfrak{SU}(3)$.

In the $\Xi $ case, the reconstruction is rooted in an \So3 symmetry
of the physically available evolution operator; this symmetry
provides information about a single subspace. In the one-atom case,
the Hilbert space contains precisely a single \So3 subspace, so the
density matrix can be completely reconstructed. In the multiple-atom
case, only reconstruction in one pre-determined subspace is
possible. In this case, our protocol would be to apply the sequence
of pulses of Eq.(\ref{Sigmapulsesequence}) with a subsequent
measurement of the number of $2A$ of photons in the irradiated
field, giving us the tomogram appearing in the reconstruction
formula of Eq.(\ref{rho_sigma}).  For completeness, we note here
that we did not consider the effective two-photon-like transition in
the $\Xi $ system due to extremely narrow width of such transitions
($\sim g^{2}/\Delta $), which leads to serious experimental
difficulties in its detection.

In the case of $\Lambda $ configuration, the evolution operator
generates an $\mathfrak{SU}(2)$ transformation and, even in the
one--atom case, there is always more than a single
$\mathfrak{SU}(2)$ multiplet: a complete reconstruction is
impossible because there always exists invariant $\mathfrak{SU}(2)$
"dark" subspaces, which cannot be uniquely identified by measuring
irradiated photons. We stress that the decomposition of the Hilbert
space into invariant subspaces occurs as a result of the inability
to access independent transitions separately;  this to be contrasted
with the approach of Ref.\cite{Karassiov}, wherein
$\mathfrak{SU}(2)$ decomposability arises from considerations of
perfectly general polarization states.  Note also that although the
effective transitions between degenerate levels in the $\Lambda $
case are not sensitive to the atom-field detunings, they still
require long interaction times.

The tomographic protocol for
$\Lambda$ differs from the $\Xi$. After application of the sequence
of pulses Eq.(\ref{ROTADA}) to a
$\Lambda$--type atom, we have to measure the probability of
detecting zero irradiated photons, which leads to the tomogram used
in Eq.(\ref{reconsfornondeg}).

Finally, we observe that the tomographic reconstruction process for
a collection of non-degenerate three-level atoms is a simple
generalization of the familiar process used for two-level quantum
systems.  In both instances, one uses the whole dynamic symmetry
group to carry out the inversion process. In contrast to this, we
are restricted to a specific subgroup in the degenerate cases, which
essentially reduces our tools and actually limits the possibility of
the complete tomographic reconstruction.

We would like to thank Dr. O. Aguilar for his participation in the early
stages of this project. The work of A. B. Klimov is partially supported by
grants CONACyT 45704. The work of H. de Guise is supported by NSERC of
Canada.

\appendix

\section{$\mathfrak{SU}(3 )$ basis states and $\mathcal{D}$-functions}

\label{Dfunctiondetails}

In this section, we review some notation useful mostly in section \ref%
{nondegenerate}. Further details can be found in \cite{Hubert}.

The correspondence between the occupational basis and states of the $(A,0)$
is
\begin{equation}
\vert n_1n_2n_3 \rangle\mapsto\vert (A,0)nI \rangle\, .
\label{su3stateappenda}
\end{equation}
In Eq.(\ref{su3stateappenda}) and throughout this paper, $n$ is a
shorthand for
$(n_1n_2n_3)$. Here, the
$(A,0)$ labels indicate that $\vert n_1n_2n_3 \rangle$ can be
reached, using the $\hat S_{ij}$ operators of Eq.(\ref{Schwinger}),
from the
state $\vert A00 \rangle$. This state is killed by the so--called $\mathfrak{%
su}(3 )$ raising operators $\hat S_{12},\hat S_{13}$ and $\hat S_{23}$. The
eigenvalues of the $\mathfrak{su}(3 )$ diagonal operators
\begin{equation}
\hat\mathfrak{h}_1=\hat S_{11}-\hat S_{22}\, ,\qquad \hat\mathfrak{h}_2=\hat
S_{22}-\hat S_{33}\, ,
\end{equation}
acting on $\vert A00 \rangle$ are, respectively, $(A,0)$.

The angular momentum label $I$ is necessary to deal with the general case
considered in Ref.\cite{Hubert}, where states of more general families of
the type $(p,q)$ are constructed. A state $\vert (p,q)nI \rangle$ can be
reached from the state $\vert (p,q)(p+q,q,0)\textstyle{\frac{1}{2}} p \rangle
$, \emph{i.e.} with $n_1=p+q,n_2=q,n_3=0$ and $I=\textstyle{\frac{1}{2}} p$.
$\vert (p,q)(p+q,q,0)\textstyle{\frac{1}{2}} p \rangle$ is killed by the $%
\mathfrak{su}(3 )$ raising operators, and the eigenvalues of $(\hat\mathfrak{%
h}_1,\hat\mathfrak{h}_2)$ are $(p,q)$. When $p$ and $q$ are both
non--zero, it is possible to have distinct states in the same
$(p,q)$ family that have identical $n$, so the angular momentum
label $I$ is required to distinguish these distinct states.

Some calculations require the evaluation of the matrix elements
\begin{eqnarray}
\langle n_1n_2n_3 \vert\bar U(\sigma)\vert n_1n_2n_3 \rangle^*&=& \langle
(A,0)nI \vert\bar U(\sigma) \vert (A,0)\nu I^{\prime}\rangle^*\, ,  \nonumber
\\
&\equiv&\mathcal{D}^{(A,0)*}_{n I,\nu I^{\prime}}(\sigma)\, .
\end{eqnarray}
This matrix element is related to the matrix element between basis states of
the irrep $(0,A)$, which is conjugate to $(A,0)$, by
\begin{equation}
\mathcal{D}^{(A,0)*}_{n I,\nu I^{\prime}}(\tau)=(-1)^{n_2+\nu_2}\, \mathcal{D%
}^{(0,A)}_{n^* I,\nu^* I^{\prime}}(\tau)\, .  \label{Dfunctionstar}
\end{equation}
Here,
\begin{equation}
\mathcal{D}^{(0,A)}_{n^* I,\nu^* I^{\prime}}(\tau)\equiv \langle (0,A)n^* I
\vert\bar U(\sigma)\vert (0,A)\nu^* I^{\prime}\rangle\, .
\end{equation}
The dimension of $(0,A)$ is the same as the dimension of $(A,0)$,
but the construction of Ref.\cite{Hubert} for basis state of $\vert
(0,A)n^* I \rangle
$ requires twice as many quanta as the basis states of $\vert (A,0) n I
\rangle $. The relation between $n$ and $n^*$ is
\begin{equation}
n=(n_1,n_2,n_3)\mapsto n^*=(A-n_1,A-n_2,A-n_3)\, .
\end{equation}
Using this and the results from \cite{Hubert}, one can verify Eq.(\ref%
{Dfunctionstar}).

Using Eq.(\ref{Dfunctionstar}), one can also verify that the $\mathcal{D}$
functions are orthogonal, in the sense that
\begin{equation}
\frac{\hbox{\rm dim}(\lambda,\mu)}{1024 \pi^5}\,\displaystyle\int\,d\Omega\,%
\mathcal{D}^{(\lambda,\mu)*}_{nI,\nu L}(\Omega) \mathcal{D}^{(\lambda^\prime
,\mu^\prime )}_{n^\prime I^\prime,\nu^\prime L^\prime}(\Omega)
=\delta_{\lambda\lambda^\prime }\delta_{\mu\mu^\prime
}\delta_{nn^\prime}\delta_{II^\prime}\delta_{\nu\nu^{\prime}}\delta_{LL^%
\prime}\, ,
\end{equation}
where
\begin{equation}
d\Omega=\sin\beta_1\,\cos\textstyle{\frac{1}{2}}\beta_2 \left(\sin\textstyle{%
\frac{1}{2}}\beta_2\right)^3\,\sin\beta_3\,
\,d\alpha_1\,d\beta_1\,d\gamma_1\,d\alpha_2\,d\beta_2\,d\alpha_3\,d\beta_3\,
d\gamma_3  \label{measure}
\end{equation}
is the invariant measure, which can be found in the usual ways \cite%
{invariantmeasure}. The normalization follows from the dimensionality
formula
\begin{equation}
\hbox{\rm dim}(\lambda,\mu)=\textstyle{\frac{1}{2}}(\lambda+1)(\mu+1)(%
\lambda+\mu+2)
\end{equation}
for the irrep $(\lambda,\mu)$ and the use the parameter range
\begin{equation}
\begin{array}{ccccccccccccccccc}
0 & \le & \alpha_1 & \le & 4\pi\, , & \quad & 0 & \le & \beta_1 & \le &
\pi\, , & \quad & 0 & \le & \gamma_1 & \le & 4\pi\, , \\
0 & \le & \alpha_2 & \le & 2\pi\, , &  & 0 & \le & \beta_2 & \le & \pi\, , &
&  &  &  &  &  \\
0 & \le & \alpha_3 & \le & 4\pi\, , &  & 0 & \le & \beta_3 & \le & \pi\, , &
& 0 & \le & \gamma_3 & \le & 4\pi\, .%
\end{array}%
\end{equation}

\section{Reduced $\mathfrak{SU}(3 )$ Clebsch-Gordan coefficients}
\label{su3CGcoeffs}

\subsection{Basis states}

The construction of states in the irrep $(p,q)$ of $\mathfrak{su}(3 )$ is
detailed in \cite{Hubert}. We can summarize this procedure by stating that
one requires, at a minimum, a total of $p+2q$ bosons. These bosons must be
of at least two types when $q\ne 0$. Thus, if $a_{ij}^{\dagger}$ creates a
boson of type $j$ in mode $i$, we define, quite generally,
\begin{equation}
\hat S_{k\ell}=a_{k1}^{\dagger}a_{\ell 1}+a_{k2}^{\dagger}a_{\ell 2}\, .
\end{equation}

If $\vert 0 \rangle$ denotes state with no boson excitation, the state
\begin{equation}
\left\vert%
\begin{array}{cc}
a_{11}^{\dagger} & a_{12}^{\dagger} \\
a_{21}^{\dagger} & a_{22}^{\dagger}%
\end{array}%
\right\vert^q\,(a_{11}^{\dagger})^p\, \vert 0 \rangle \sim \vert
(p,q)(p+q,q,0);\textstyle{\frac{1}{2}} q \rangle\, ,  \label{su3pq}
\end{equation}
containing $p+q$ bosons in mode 1, $q$ boson in mode $2$ and none in mode $3$%
, belongs to the $(p,q)$ irrep. It is, in fact, killed by every $\hat
C_{k\ell}$ with $\ell>k$ and is thus the highest weight state of $(p,q)$.
Here,
\begin{equation}
\left\vert%
\begin{array}{cc}
a_{11}^{\dagger} & a_{12}^{\dagger} \\
a_{21}^{\dagger} & a_{22}^{\dagger}%
\end{array}%
\right\vert =a_{11}^{\dagger}a_{22}^{\dagger}-a_{12}^\dagger a_{21}^\dagger
\end{equation}
is the determinant of the matrix. Other states in $(p,q)$ are obtained by
laddering down from $\vert (p,q)(p+q,q,0);\textstyle{\frac{1}{2}} q \rangle$.

This is not the only possibility. One can verify that
\begin{eqnarray}
&&\vert (p,q)(p+q+t,q+t,t);\textstyle{\frac{1}{2}} q )  \nonumber \\
&&\qquad\qquad =\left\vert%
\begin{array}{ccc}
a_{11}^{\dagger} & a_{12}^{\dagger} & a_{13}^\dagger \\
a_{21}^{\dagger} & a_{22}^{\dagger} & a_{23}^\dagger \\
a_{31}^\dagger & a_{32}^\dagger & a_{33}^{\dagger}%
\end{array}%
\right\vert^t\, \left\vert%
\begin{array}{cc}
a_{11}^{\dagger} & a_{12}^{\dagger} \\
a_{21}^{\dagger} & a_{22}^{\dagger}%
\end{array}%
\right\vert^q\, (a_{11}^{\dagger})^p\, \vert 0 \rangle \, ,  \nonumber \\
&&\qquad\qquad \sim \left\vert%
\begin{array}{ccc}
a_{11}^{\dagger} & a_{12}^{\dagger} & a_{13}^\dagger \\
a_{21}^{\dagger} & a_{22}^{\dagger} & a_{23}^\dagger \\
a_{31}^\dagger & a_{32}^\dagger & a_{33}^{\dagger}%
\end{array}%
\right\vert^t\, \vert (p,q)(p+q,q,0);\textstyle{\frac{1}{2}} q \rangle\, ,
\end{eqnarray}
visibly contains $p+2q+3t$ bosons but is equivalent to the state of Eq.(\ref%
{su3pq}) because the determinant
\begin{equation}
\left\vert%
\begin{array}{ccc}
a_{11}^{\dagger} & a_{12}^{\dagger} & a_{13}^\dagger \\
a_{21}^{\dagger} & a_{22}^{\dagger} & a_{23}^\dagger \\
a_{31}^\dagger & a_{32}^\dagger & a_{33}^{\dagger}%
\end{array}%
\right\vert
\end{equation}
is an $\mathfrak{SU}(3 )$ scalar.

More generally, if the usual ket $\vert (p,q)nI \rangle$ denotes a state in $%
(p,q)$ containing $p+2q$ bosons, then the (round) ket
\begin{equation}
\vert (p,q)nI )= \left\vert%
\begin{array}{ccc}
a_{11}^{\dagger} & a_{12}^{\dagger} & a_{13}^\dagger \\
a_{21}^{\dagger} & a_{22}^{\dagger} & a_{23}^\dagger \\
a_{31}^\dagger & a_{32}^\dagger & a_{33}^{\dagger}%
\end{array}%
\right\vert^t\,\vert (p,q)nI \rangle
\end{equation}
differs from $\vert (p,q)nI \rangle$ by at most a normalization but contains
$p+2q+3t$ bosons.

\subsection{$\mathfrak{SU}(3 )$ Clebsch-Gordan coefficients}

The $\mathfrak{SU}(3)$-coupling $(A,0)\otimes(0,A)$ can be
decomposed in the direct sum \cite{su3tensor}
\begin{eqnarray}
(A,0)\otimes(0,A)&=&(A,A)\oplus(A-1,A-1)\oplus\ldots\oplus (0,0)\, ,
\nonumber \\
&=&\bigoplus_{\lambda=0}^{A}\,(\lambda,\lambda)\, .  \label{su3series}
\end{eqnarray}
The irrep $(\sigma,\sigma)$ occurs at most once in the decomposition.

To compute $\mathfrak{SU}(3 )$ Clebsch-Gordan coefficients for states in the
series of Eq.(\ref{su3series}), we must couple states of the form
\begin{equation}
\vert (A,0)nI_1 \rangle\vert (0,A)\nu^*I_2 \rangle\, ,
\end{equation}
which contain a total of $3A$ bosons of three types. States in the irrep $%
(\sigma,\sigma)$ of the series of Eq.(\ref{su3series}) are of the form
\begin{equation}
\vert (\sigma,\sigma)NI_3 )= \left\vert%
\begin{array}{ccc}
a_{11}^{\dagger} & a_{12}^{\dagger} & a_{13}^\dagger \\
a_{21}^{\dagger} & a_{22}^{\dagger} & a_{23}^\dagger \\
a_{31}^\dagger & a_{32}^\dagger & a_{33}^{\dagger}%
\end{array}%
\right\vert^{A-\sigma}\, \vert (\sigma,\sigma)N^{(\sigma)}I_3 \rangle\, ,
\end{equation}
where $\vert (\sigma,\sigma)N^{(\sigma)}I_3 \rangle$ is the state with $%
3\sigma$ bosons described in \cite{Hubert}.

Note that, because the state $\vert (\sigma,\sigma)N^{(\sigma)}I_3 \rangle$
does not contain $3A$ bosons, we do not have $n_i+\nu_i^*=N^{(\sigma)}_i$
\emph{etc} but rather
\begin{equation}
n_i+\nu_i^*=N^{(\sigma)}_i+(A-\sigma)\, .  \label{Nsigmadef}
\end{equation}

Thus we have
\begin{eqnarray}
\vert (\sigma,\sigma)N^{(\sigma)}I_3 \rangle&=&\sum_{n\nu I_1I_2} \vert
(A,0)nI_1 \rangle\vert (0,A)\nu^* I_2 \rangle  \nonumber \\
&&\qquad\times\sucg{(A,0)}{nI_1}{(0,A)}{\nu^*I_2}{(\sigma,\sigma)}{N^{(\sigma)}I_3}
\, ,\label{c11}
\end{eqnarray}
where $M_1=\textstyle{\frac{1}{2}}(n_2-n_3)\, ,M_2=\textstyle{\frac{1}{2}}%
(\nu_3-\nu_2)$. The phases of the states $\vert (A,0)nI_1 \rangle\,
,\vert (0,A)\nu^*I_2 \rangle$ and $\vert
(\sigma,\sigma)N^{(\sigma)}I_3 \rangle$ are those of
Ref.\cite{Hubert}. The phase of the Clebsch-Gordan coefficient is
determined by forcing
\begin{equation}
\hbox{\rm sign}\left(\sucg{(A,0)}{(A00)\textstyle{\frac{1}{2}}
A}{(0,A)}{\nu^*I_2}{(\sigma,\sigma)}{(2\sigma,\sigma,0)\textstyle{\frac{1}{2}}
\sigma}  \right)=+\, .
\end{equation}

As always, it is convenient to rewrite Eq.(\ref{c11}) as
\begin{eqnarray}
\vert (\sigma,\sigma)N^{(\sigma)}I_3 \rangle &=&\sum_{n\nu} \vert (A,0)nI_1
\rangle\vert (0,A)\nu^* I_2 \rangle\,   \nonumber \\
&&\times
\surcg{(A,0)}{n_1I_1}{(0,A)}{\nu_1^*I_2}{(\sigma,\sigma)}{N^{(\sigma)}_1I_3}
C^{I_3\,M_3}_{I_1\,M_1,\, I_2\,M_2}\,
\end{eqnarray}
where $C^{I_3\,M_3}_{I_1\,M_1,\, I_2\,M_2}$ is the usual
$\mathfrak{su}(2 )$ coupling coefficient and the reduced
Clebsch-Gordan $ \surcg{(A,0)}{n_1I_1}{(0,A)}{\nu_2^*I_2}{(\sigma,\sigma)}{N^{(\sigma)}_1I_3}$ does not depend on $M_i$%
. Tables are provided for $A=2$ and $A=1$.

\begin{table}[tbp]
\caption{1-atom case. Reduced $\mathfrak{SU}(3 )$ Clesbsh-Gordan
coefficients for $(1,0)\otimes(0,1)\to(0,0)$.}
\begin{center}
\begin{tabular}{|c|c||c|c||c|c||c|}
\hline $\ N_1\ $ & $\ I_3\ $ & $\ n_1\ $ & $\ I_1\ $ & $\ \nu_1^*\ $
& $\ I_2 \ $ & $\surcg{(1,0)}{n_1I_1}{(0,1)}{\nu_1^*I_2}{(0,0)}{N^{(\sigma)}_1I_3}$ \\
\hline\hline
0 & 0 & 0 & $\textstyle{\frac{1}{2}}$ & 1 & $\textstyle{\frac{1}{2}}$ & $-%
\sqrt{\frac{2}{3}}$ \\ \hline
&  & 1 & 0 & 0 & 0 & $+\sqrt{\frac{1}{3}}$ \\ \hline
\end{tabular}%
\end{center}
\end{table}

\begin{table}[tbp]
\caption{1-atom case. Reduced $\mathfrak{SU}(3 )$ Clesbsh-Gordan
coefficients for $(1,0)\otimes(0,1)\to(1,1)$.}
\begin{center}
\begin{tabular}{|c|c||c|c||c|c||c|}
\hline
$\ N_1\ $ & $\ I_3\ $ & $\ n_1\ $ & $\ I_1\ $ & $\ \nu_1^*\ $ & $\ I_2 \ $ &
$\surcg{(1,0)}{n_1I_1}{(0,1)}{\nu_1^*I_2}{(1,1)}{N^{(\sigma)}_1I_3} $ \\
\hline\hline 2 & $\textstyle{\frac{1}{2}}$ & 1 & 0 & 1 &
$\textstyle{\frac{1}{2}}$ & $+1$
\\ \hline\hline
1 & 1 & 0 & $\textstyle{\frac{1}{2}}$ & 1 & $\textstyle{\frac{1}{2}}$ & $+1$
\\ \hline\hline
1 & 0 & 0 & $\textstyle{\frac{1}{2}}$ & 1 & $\textstyle{\frac{1}{2}}$ & $+%
\sqrt{\frac{1}{3}}$ \\ \hline
&  & 1 & 0 & 0 & 0 & $+\sqrt{\frac{2}{3}}$ \\ \hline\hline
0 & $\textstyle{\frac{1}{2}}$ & 0 & $\textstyle{\frac{1}{2}}$ & 0 & 0 & +1
\\ \hline
\end{tabular}%
\end{center}
\end{table}

\begin{table}[tbp]
\caption{2-atom case. Reduced $\mathfrak{SU}(3 )$ Clesbsh-Gordan
coefficients for $(2,0)\otimes(0,2)\to(0,0)$.}
\begin{center}
\begin{tabular}{|c|c||c|c||c|c||c|}
\hline
$\ N_1\ $ & $\ I_3\ $ & $\ n_1\ $ & $\ I_1\ $ & $\ \nu_1^*\ $ & $\ I_2 \ $ &
$\surcg{(2,0)}{n_1I_1}{(0,2)}{\nu_1^*I_2}{(0,0)}{N^{(\sigma)}_1I_3}$ \\
\hline 0 & 0 & 0 & 1 & 2 & 1 & $+\sqrt{\frac{1}{2}}$ \\ \hline
&  & 1 & $\textstyle{\frac{1}{2}}$ & 1 & $\textstyle{\frac{1}{2}}$ & $-\sqrt{%
\frac{1}{3}}$ \\ \hline
&  & 2 & 0 & 0 & 0 & $+\sqrt{\frac{1}{6}}$ \\ \hline
\end{tabular}%
\end{center}
\end{table}

\begin{table}[tbp]
\caption{2-atom case. Reduced $\mathfrak{SU}(3 )$ Clesbsh-Gordan
coefficients for $(2,0)\otimes(0,2)\to(1,1)$.}
\begin{center}
\begin{tabular}{|c|c|c|c|c|c||c|}
\hline
$\ N_1\ $ & $\ I_3\ $ & $\ n_1\ $ & $\ I_1\ $ & $\ \nu_1^*\ $ & $\ I_2\ $ &
$\surcg{(2,0)}{n_1I_1}{(0,2)}{\nu_1^*I_2}{(1,1)}{N^{(\sigma)}_1I_3}$ \\
\hline\hline
2 & $\textstyle{\frac{1}{2}}$ & 1 & $\textstyle{\frac{1}{2}}$ & 2 & 1 & $-%
\sqrt{\frac{3}{5}}$ \\ \hline
&  & 2 & 0 & 1 & $\frac{1}{2}$ & $+\sqrt{\frac{2}{5}}$ \\ \hline\hline
1 & 0 & 0 & 1 & 2 & 1 & $-\sqrt{\frac{2}{5}}$ \\ \hline
&  & 1 & $\textstyle{\frac{1}{2}}$ & 1 & $\textstyle{\frac{1}{2}}$ & $-\sqrt{%
\frac{1}{15}}$ \\ \hline
&  & 2 & 0 & 0 & 0 & $\sqrt{\frac{8}{15}}$ \\ \hline\hline
1 & 1 & 0 & 1 & 2 & 1 & $-\sqrt{\frac{4}{5}}$ \\ \hline
&  & 1 & $\textstyle{\frac{1}{2}}$ & 1 & $\textstyle{\frac{1}{2}}$ & $+\sqrt{%
\frac{1}{5}}$ \\ \hline\hline
0 & $\textstyle{\frac{1}{2}}$ & 0 & 1 & 1 & $\textstyle{\frac{1}{2}}$ & $-%
\sqrt{\frac{3}{5}}$ \\ \hline
&  & 1 & $\textstyle{\frac{1}{2}}$ & 0 & 0 & $+\sqrt{\frac{2}{5}}$ \\ \hline
\end{tabular}%
\end{center}
\end{table}

\begin{table}[tbp]
\caption{2-atom case. Reduced $\mathfrak{SU}(3 )$ Clesbsh-Gordan
coefficients for $(2,0)\otimes(0,2)\to(2,2)$.}
\begin{center}
\begin{tabular}{|c|c||c|c||c|c||c|c||c|}
\hline \ $N_1\ $ & \ $I_3 \ $ & $\ n_1\ $ & $\ I_1\ $ & $\ \nu_1^*\
$ & $\ I_2\ $\ &
$\surcg{(2,0)}{n_1I_1}{(0,2)}{\nu_1^*I_2}{(2,2)}{N^{(\sigma)}_1I_3}$
&  &
\\ \hline\hline
0 & 1 & 0 & 1 & 0 & 0 & +1 &  &  \\ \hline\hline
1 & $\textstyle{\frac{1}{2}}$ & 0 & 1 & 1 & $\frac{1}{2}$ & $+ \sqrt{\frac{2%
}{5}}$ &  &  \\ \hline
&  & 1 & $\textstyle{\frac{1}{2}}$ & 0 & 0 & $+\sqrt{\frac{3}{5}}$ &  &  \\
\hline\hline
1 & $\frac{3}{2}$ & 0 & 1 & 1 & $\frac{1}{2}$ & + 1 &  &  \\ \hline\hline
2 & 2 & 0 & 1 & 2 & 1 & +1 &  &  \\ \hline\hline
2 & 1 & 0 & 1 & 2 & 1 & $+\sqrt{\frac{1}{5}} $ &  &  \\ \hline
&  & 1 & $\textstyle{\frac{1}{2}}$ & 1 & $\textstyle{\frac{1}{2}}$ & $+\sqrt{%
\frac{4}{5}}$ &  &  \\ \hline\hline
2 & 0 & 0 & 1 & 2 & 1 & $+\sqrt{\frac{1}{10}}$ &  &  \\ \hline
&  & 1 & $\textstyle{\frac{1}{2}}$ & 1 & $\textstyle{\frac{1}{2}}$ & $+
\sqrt{\frac{3}{5}}$ &  &  \\ \hline
&  & 2 & 0 & 0 & 0 & $+\sqrt{\frac{3}{10}}$ &  &  \\ \hline\hline
3 & $\textstyle{\frac{1}{2}}$ & 1 & $\textstyle{\frac{1}{2}}$ & 2 & 1 & $+%
\sqrt{\frac{2}{5}}$ &  &  \\ \hline
&  & 2 & 0 & 1 & $\textstyle{\frac{1}{2}}$ & $+\sqrt{\frac{3}{5}}$ &  &  \\
\hline\hline
3 & $\frac{3}{2}$ & 1 & $\textstyle{\frac{1}{2}}$ & 2 & 1 & +1 &  &  \\
\hline\hline
4 & 1 & 2 & 0 & 2 & 1 & +1 &  &  \\ \hline
\end{tabular}%
\end{center}
\end{table}

\section{Final form of the density matrix for the non--degenerate case}
\label{sec:casenondegenerate}

In this section we present the technical steps to obtain Eq.(\ref%
{reconsfornondeg}) from Eq.(\ref{eqnondgtomo}).

First, multiply both sides of Eq.(\ref{eqnondgtomo}) by $\mathcal{D}%
^{(\mu,\mu)*}_{(\mu\mu\mu)0,\,N^{(\mu)} J^{\prime}}(\tau)$ for fixed $%
N^{(\mu)}$ and fixed $J^{\prime}$, integrate over the $\mathfrak{SU}(3)$%
--invariant measure of Eq.(\ref{measure}) and rearrange. This
produces
\begin{eqnarray}
&&\displaystyle\frac{(\mu+1)^3}{1024\pi^5}\,
\sucg{(A,0)}{(A00)0}{(0,A)}{(0AA)0}{(\mu,\mu)}{(\mu\mu\mu)0}^{-1}\, \int d\tau\, \mathcal{D}%
^{(\mu,\mu)*}_{(\mu\mu\mu)0,\,N^{(\mu)}
J^{\prime}}(\tau)\omega(\tau)
\nonumber \\
&&\qquad\qquad =\displaystyle\sum_{n\nu}\,(-1)^{\nu_2}\,
\rho_{n,\,\nu}\, \sucg{(A,0)
}{nI}{(0,A)}{\nu^*I^\prime}{(\mu,\mu)}{N^{(\mu)} J^{\prime}}\, .
\end{eqnarray}
In this last expression, the sums over $n$ and $\nu$ are not
independent but linked by Eq.(\ref{Nlambdadef}).

To complete the inversion, we use orthogonality of $\mathfrak{SU}(3
)$ CGs:
\begin{eqnarray}
&&\sum_{\mu
J^{\prime}}\sucg{(A,0)}{nI}{(0,A)}{\nu^*I^\prime}{(\mu,\mu)}{N^{(\mu)}
J^{\prime}}\,
\sucg{(A,0)}{\bar{n}\bar{I}}{(0,A)}{\bar{\nu}^*\bar{I}^\prime}{(\mu,\mu)}{N^{(\mu)} J^{\prime}}  \nonumber \\
&&\qquad\qquad =\delta_{\bar{n}n}\,\delta_{\bar I
I}\,\delta_{\bar\nu \nu}\delta_{\bar{I}^{\prime}I^{\prime}}\, ,
\end{eqnarray}
and rearrange the notation to finally yield
Eq.(\ref{reconsfornondeg}).

\end{document}